# Daubechies wavelet coefficients: a tool to study interplanetary magnetic field fluctuations

Arian Ojeda González[a,c,*] , Odim Mendes Junior[a], Margarete Oliveira Domingues.[b] , and Varlei Everton Menconi. [a]

[a]*DGE/CEA/National Institute for Space Research - INPE 12227-010 São José dos Campos, SP, Brazil*

[b]*LAC/CTE/National Institute for Space Research - INPE 12227-010 São José dos Campos, SP, Brazil*

[c]*Department of Space Geophysics, Institute of Geophysics and Astronomy - IGA Havana City, Cuba*

[*]***e-mail: ojeda.gonzalez.a@gmail.com***

**Abstract**

We have studied a set of 41 magnetic clouds (MCs) measured by the ACE spacecraft, using the discrete orthogonal wavelet transform (Daubechies wavelet of order two) in three regions: Pre-MC (plasma sheath), MC and Post-MC. We have used data from the IMF GSM-components with time resolution of 16 s. The mathematical property chosen was the statistical mean of the wavelet coefficients ($\langle Dd1 \rangle$). The Daubechies wavelet coefficients have been used because they represent the local regularity present in the signal being studied. The results reproduced the well-known fact that the dynamics of the sheath region is more than that of the MC region. This technique could be useful to help a specialist to find events boundaries when working with IMF datasets, i.e., a best form to visualize the data. The wavelet coefficients have the advantage of helping to find some shocks that are not easy to see in the IMF data by simple visual inspection. We can learn that fluctuations are not low in all MCs, in some cases waves can penetrate from the sheath to the MC. This methodology has not yet been tested to identify some specific fluctuation patterns at IMF for any other geoeffective interplanetary events, such as Co-rotating Interaction Regions (CIRs), Heliospheric Current Sheet (HCS) or ICMEs without MC signatures. In our opinion, as is the first time that this technique is applied to the IMF data with this purpose, the presentation of this approach for the Space Physics Community is one of the contributions of this work.

**Resumen**

Hemos estudiado un conjunto de 41 nubes magnéticas (MCs) detectadas por el satélite ACE, utilizamos la transformada wavelet ortogonal discreta (usando wavelet de Daubechies de orden dos) en tres regiones: vaina de plasma, nube y posterior a la nube. Trabajamos con datos de las componentes del campo magnético interplanetario (IMF) en el sistema de coordenadas GSM con resolución temporal de 16 s. Se ha elegido como herramienta matemática la media estadística de los coeficientes wavelets ($\langle Dd1 \rangle$). Los coeficientes wavelets de Daubechies se han utilizado porque ellos representan la regularidad local presente en la señal de estudio. Los resultados reprodujeron el hecho bien conocido, que la dinámica es mas compleja en la vaina de plasma que en la región de la MC. Esta técnica podría ser útil a un especialista en ayudarlo encontrar fronteras de eventos cuando se trabaja con el IMF, es decir, una mejor forma de visualizar los datos. Los coeficientes wavelets tienen la ventaja de facilitar encontrar algunos choques que serían difíciles de detectar por simple inspección visual del IMF. Podemos aprender que las fluctuaciones no son igualmente pequeñas en todas las nubes, en algunos casos las ondas pueden penetrar desde la vaina hasta la MC. Esta metodología aún no ha sido testada para identificar patrones específicos de fluctuaciones en el IMF de otros eventos interplanetarios geoefectivos, tales como, regiones de interacción corrotante (CIRs), lámina de corriente heliosférica (HCS) o para ICMEs sin características de MC. Como es la primera vez que esta técnica se aplica a los datos del IMF, opinamos que una de las contribuciones de este trabajo es la presentación de este enfoque a la Comunidad de Físicos Espaciales.

*Palabras Clave:* Electrodinámica Espacial, Nubes Magnéticas, Análisis de Series Temporales, Transformada Wavelet Discreta, Clima Espacial.

# 1. Introduction

One of the very important phenomena in space is the Interplanetary Coronal Mass Ejection (ICME) as a disturbance in the solar wind (SW) that presents a large importance due to its potential geoeffectivity. Physically, a subset of ICMEs has simple flux rope-like magnetic fields, in which, briefly, the magnetic field strength is higher than the average, the magnetic field direction rotates smoothly through a large angle, and the proton temperature is low (Burlaga et al., 1981; Klein and Burlaga, 1982; Gosling, 1990). Such events, named magnetic clouds (MCs), have received considerable attention, because they are an important source of southward interplanetary magnetic field (e.g. NS, SN and S polarity, where N ≡ north and S ≡ south).

Investigations on the relation between MCs and geomagnetic storms have been carried out by many researchers (for instance, Burlaga et al., 1981; Klein and Burlaga, 1982; Gonzalez and Tsurutani, 1987; Tsurutani et al., 1988; Tsurutani and Gonzalez, 1992; Farrugia et al., 1995; Lepping et al., 2000; Dal Lago, et al., 2000; Dal Lago et al., 2001; Wu and Lepping, 2002a,b) with many purposes. Echer et al. (2005) studied a total of 149 MCs from 1966 to 2001, where 51 are of the NS type, 83 of the type SN, and 15 unipolar (N or S). They did a statistical study of MC parameters and geoeffectiveness that was determined by classifying the number of MCs followed by intense, moderate and weak magnetic storms, and by calm periods. They found that around 77% of the MCs present geoeffectivity with Dst ≤ -50 nT. Taking into account weak storms (-50 nT ≤ Dst ≤ -30 nT), 97% of MCs were followed by geomagnetic activity.

Another significant example is the work of Huttunen et al. (2005), where they studied the geomagnetic response of MCs using the 1-h Dst index. They focused on whether the storm was caused by sheath fields or by the MC itself. They found that the geomagnetic response of a MC depends greatly on its flux-rope type.

Inside ICMEs, the measured plasma velocity typically has a linear variation along the spacecraft trajectory. A much higher velocity is present in the front than in the rear, indicating expansion (Démoulin and Dasso, 2009). Burlaga and Behannon (1982) found consistency between the expansion speed estimated from in situ observations and the increase of their typical size, obtained from measurements with different spacecraft located between 2 and 4 AUs.

The MCs closer to the Sun, i.e., the ones that are near 1 AU, had higher plasma densities than the ones surrounding SW. The density inside the flux tubes has a rapid decrease with the increasing distance from the Sun where the cloud undergoes a radial expansion. The density in MCs is generally higher than average fast SW, and the slow SW, at close distances to the Sun. Bothmer and Schwenn (1998) observed that MCs in which the densities are found to be considerably lower compared to those of the ambient slow SW should have undergone strong expansion on their way out from the Sun.

Typically, the MC magnetic field configuration may be described by a force-free model as a simple approximation useful in interpreting time series data (e.g., Lundquist, 1950; Lepping et al., 1990; Burlaga, 1988; Osherovich and Burlaga, 1997; Lepping et al., 1997; Burlaga, 1995; Bothmer and Schwenn, 1998; Dasso et al., 2005). Three characteristic speeds are derived from MHD theory; these are the sound speed, the Alfvén speed, and the magnetoacoustic speed. Then five kinds of MHD shocks (fast shock, slow shock and three kinds of intermediate shocks) can be found (Burlaga, 1995, p.70). In SW have been studied the fast shock and slow shock. The magnetic field strength increases across a fast shock and decreases across a slow shock (Burlaga, 1995, p.70). A shock moving away from the Sun relative to the ambient medium is called a "forward shock". A shock moving toward the Sun relative to the ambient medium is called "reverse shock" (Gosling, 1998). In MHD, the shocks are further classified on the basis of the angle between $n$ and the ambient magnetic field observation $B$. Therefore, shocks are classified as perpendicular, parallel and oblique. The sheath is the turbulent region between a shock and an MC (Burlaga, 1995, p.132). The SW form sheaths around solar system objects: the heliosheath around the heliosphere, cometosheaths around comets and ICME-sheaths around fast ICMEs, etc. Siscoe and Odstrcil (2008) defined two types of sheath, "propagation sheath" and "expansion sheath", but pure expansion sheaths are less common than propagation sheaths. The studies on the dynamics of those kinds of electrodynamics structures are among the current concerns of the space community.

Other studies also suggest that the Interplanetary Magnetic Field (IMF) fluctuations can be geo-effective, and then the reason for space weather studies on variability related to the interplanetary phenomena (Lyons et al., 2009). According to Lyons et al. (2009) and Kim et al. (2009), the interplanetary ULF fluctuations are an important contributor to the large-scale transfer of SW energy to the magnetosphere-ionosphere system, and to the occurrence of disturbances such as substorms. In their work, the data are processed using a fast Fourier transform algorithm with 128 points (2 h) moving window to produce the power spectral density in the ULF Pc5 frequency range. Kim et al. (2009) show dynamic spectrograms of the IMF $B_z$ obtained from 1-min-resolution time-shifted ACE data for the four different SW conditions that was examined. Borovsky (2012) studied the plasma fluctuations in a dataset measured by the ACE spacecraft. All of them are using Fourier transform algorithms in a skilled way.

However, some complicated fluctuations in SW plasma could be investigated by using techniques based on approaches from nonlinear dynamics (e.g. Ojeda et al., 2005; Ojeda et al., 2013). Thus, an interesting expectation is to study the ICMEs by the analyses of the time series of the IMF, because this field should preserve intrinsic aspects of the physical structures involved. Also, IMF data studies require analysis of random or non-deterministic time series, as well as analyses taking into account the non-stationary behaviour of data. The use of wavelet coefficients has proved to be a useful technique for study those kinds of data, specially of non-stationary time series (e.g. Mendes et al., 2005; Domingues et al., 2005).

The mathematical property chosen in this work is the statistical mean of the wavelet coefficients obtained by applying the discrete orthogonal wavelet transform using Daubechies wavelet of order two (i.e. Daubechies scale filters order 2, db2). The analysis is done using the components of the IMF as recorded by the instruments of the Magnetic Field Experiment (MAG) on board of the ACE S/C at the L1 point. Therefore, our interest is to study the wavelet coefficients behaviour for diagnose of disturbance level in interval of the SW data containing the MC occurrences. The tool feature explored here is the identification of regularity/no-regularity in a function that represents the physical process (see for example, Appendix A).

As used in this work, a methodology is presented to help the solar/heliospheric physics community efforts to deal with the MCs. The wavelet analysis has important advantages, adding resources to other classical mathematical tools that could be used to study SW fluctuations. The wavelet coefficients allow to find fluctuations with pseudo-frequencies corresponding to the scales given by j, the chosen wavelet function, and the sampling period. The idea is to associate a purely periodic signal of frequency Fc with a given wavelet. The frequency maximizing the Fourier transform of the wavelet function is the central frequency (Fc) of it. It enables plotting the wavelet with an associated approximation based on the center frequency. This center frequency captures the main wavelet oscillations. Thus, the center frequency is a convenient and simple characterization of the leading dominant frequency of the wavelet (Abry, 1997).

As we are interested in studying fluctuations with larger frequencies (in this case on data from 16-second time resolution), the Daubechies function db2 with one decomposition level seems an appropriate choice. A zooming in analysing the IMF fluctuations with a pseudo-period of 48 seconds could help to better locate the ICME boundaries. Thus, a statistical study has to be performed. For this reason, three regions from 41 ICMEs will be studied, i.e. plasma sheath, magnetic cloud, and region after the MC.

The aim of this work is to characterize the wavelet coefficients amplitudes of the magnetic field at the three different regions around an ICME event to relate it to features of the interplanetary medium. The primary idea is to distinguish more quiescent periods (in terms of magnetic variation) related to MC from non-quiescent periods of two other processes. For the use of magnetic field data, the motivation is that in many cases there are only those kinds of data available for investigation. The content is organized as follows. Section 2 presents dataset. Section 3 describes the implemented methodology. Section 4 discusses the results. Section 5 gives the conclusions.

## 2. IMF Dataset

The Lagrangean point L1 is a gravitational equilibrium point between the Sun and the Earth at about 1.5 million km from Earth and 148.5 million km from the Sun (Celletti and Giorgilli, 1990). The data used here are from Advance Composition Explorer (ACE) spacecraft, which has been making such measurements orbiting L1 since 1997 (Smith et al., 1998). From its location, ACE has a prime view of the SW, the IMF and the higher energy particles accelerated by the Sun, as well as particles accelerated in the Heliosphere and the galactic regions beyond. The plasma particles detected by ACE arrive at the magnetopause after about 30 min (Smith et al., 1998). The MAG on board ACE consists of twin vector fluxgate magnetometers to measure IMF (Smith et al., 1998). The data contains time averages of the magnetic field over time periods 1 s, 16 s, 4 min, hourly, daily and 27 days (1 Bartels rotation).

In this work we use data from the IMF GSM-components with time resolution of 16 s. We work with 41 of 80 events (73 MCs and 7 cloud candidate) identified by Huttunen et al. (2005). These events are shown in chronological order in Table 1. The columns from left to right give: a numeration of the events, year, shock time (UT), MC start time (UT), MC end time (UT), and the end time (UT) of the third region respectively.

A total of 17 events listed in Table 2 are not treated in this work. The reason is that the ACE data before about the end of 1997 were not qualified for research use. Huttunen et al. (2005) used the measurements recorded by the WIND spacecraft for this initial period. The magnetic field instrument (MFI) on board WIND is composed of dual triaxial fluxgate magnetometers. We avoid in this analysis mixing dataset from different types of spacecraft. Another problem is that the WIND data available in averages present 3 s, 1 min, and 1 h time resolution, a lower resolution than the one we used by ACE.

The MC events that are not associated with shock waves are not tested here. They are presented in Table 3. The purpose of this selection, in this exploratory study, is to deal with the cases presenting the three periods (clear Pre-MC, MC and Post-MC). Thus, with the well-defined MC cases, the assumption is to objectively unravel the magnetically quiescent interval related to the MC period. If there are significant differences of the coefficient features among the periods, then this tool can be used to identify boundaries of ICMEs in most clear basis. Other SW disturbances different of MCs are not studied here.

## 3. Methodology

The Discrete Wavelet Transform (DWT) is a linear multilevel efficient transform that is very popular in data compression (Mallat, 1989; Daubechies, 1992; Hubbard, 1997). Mathematically, this transform is built based on a multiscale tool called Multiresolution analysis $\{V^j, \Phi\} \in L^2$ proposed by S. Mallat (see details in Mallat (1989)), where $\Phi$ is a scale function, $V^j = span\{\Phi_k^j\}$, and $L^2$ is the functional space of the square-integrable functions. The DWT uses discrete values of scale $(j)$ and position $(k)$.

The great contribution of wavelet theory is the characterization of complementary spaces between two embedded spaces $V^{j+1} \subset V^j$, through direct sums $V^j = V^{j+1} + W^{j+1}$, where $W^j = span\{\Psi_k^j\}$ with $\Psi$ the wavelet function.

Mallat also developed an efficient and very simple way to compute this multilevel transform based on filter banks. With this tool, one can compute the so called discrete scale coefficient $c_k^j$ and wavelet coefficient $d_k^j$ associate with discrete values of scale $j$ and position $k$. Roughly speaking, the basic ingredients to compute one level step of this transform are the low filter ($h$) related to the analysing scale function and its relation with the high-pass filter ($g$) related to the analysing wavelet function. These filters are used to compute the scale coefficients and the wavelet coefficients as follows:

$$c_k^j = \sqrt{2} \sum h(m-2k) c_m^{j+1} \qquad (1)$$

and

$$d_k^j = \sqrt{2} \sum g(m-2k) c_m^{j+1}. \qquad (2)$$

The multilevel transform is done by repeating this procedure recursively: convolute the scale coefficients with the filter and performing the downsampling procedure, i.e., removing one data point between two. Therefore in each scale decomposition levels the number of data is reduced by two. Following is a scheme for the DWT and its inverse (IDWT),

$$\{c^{j+1}\} \underset{IDWT}{\overset{DWT}{\leftrightarrow}} \{c^j, d^j, d^{j-1} \cdots, d^{j-j_0}\}.$$

The initial data is consider as first level scale coefficient $c^{j+1}$.

The wavelet coefficients have the property that their amplitudes are related to the local regularity of the analysed data (Mallat, 1989; Daubechies, 1992). This means that, where the data has a smooth behaviour, the wavelet coefficients are smaller, and vice-versa. This is the basic idea of data compression and the application we are doing here. The wavelet coefficient amplitudes are also related to the analysing wavelet order and the scale level.

There is not a perfect wavelet choice for a certain data analysis. However, one can follow certain criteria to provide a good choice, see for instance, Domingues et al. (2005).

In this work, we have chosen the Daubechies scaling function of order 2, with the choice that the wavelet function locally reproduces a linear polynomial. On one hand, high order analysing Daubechies functions are not adding a better local reproduction of the MC disturbance data. On the other hand, the analysing function of order 1 does not reproduce well these disturbances locally.

We have also observed that just one decomposition level is enough for the energy analysis methodology that we propose here, which corresponds to a pseudo-period of 48 s. The pseudoperiod is $T_a = (a\Lambda)/F_c$ where $a = 2^j$ is a scale, $\Lambda = 16\,s$ is the sampling period, Fc = 0.6667 is the center frequency of a wavelet in Hz (Abry, 1997). In Table 4, as a test, some decomposition levels and the Daubechies scaling function of order 1 to 4 are shown, where $F_c$= [0.9961; 0.6667; 0.8000; 0.7143]. Pseudo-periods (seconds) regarding the Daubechies orthogonal wavelets are presented. It also shows that the information here could be useful for studying fluctuations with different frequencies which is not done in this work.

The non zero values of the low filter h for Daubechies order 2 analysing wavelet are:

$[h_0, h_1, h_2, h_3]$ =0.4829629131445, 0.8365163037378, *[ 0.2241438680420, - 0.1294095225512]* and $[g_0, g_1, g_2, g_3] = [h_0, -h_1, h_2, -h_3]$ is the high-pass band filter (Daubechies, 1992, p.195).

In this case, we are using an orthogonal transform. The orthogonal property is very important here, because with it we with it, we can guarantee a preserving energy property in the wavelet transform similarly to the Parseval theorem for Fourier analysis (Daubechies, 1992). Therefore the total energy of the signal is equal to the superposition of the individual contributions of energy of their wavelet coefficient in each decomposition level (Holschneider, 1991).

In the characterization of a SW disturbance, we perform one decomposition level and we compute the square of wavelet coefficients $(d^1$ or $d1)$ (energy content on that level), as in Mendes da Costa (2011); Mendes et al. (2005), and its mean value $D_{d1}$ is:

$$D_{d1} = \frac{\sum_{i=1}^{N/2} d1_i^2}{N/2}, \text{ where } N = length(f(t)). \qquad (3)$$

This value was calculated in the three regions for each IMF components ($B_x, B_y, B_z$). Its values are influenced by the fluctuations amplitude in the physical system studied. It is lower at a system in stationary state with minimum energy. If the system has a strong external perturbation then the $D_{d1}$ value increases.

The MCs have flux-rope-like topology and form a large-scale winding of a closed magnetic structure that could be nearly force-free. And it is possible to see anisotropy of magnetic field fluctuations in an average interplanetary MC at 1 AU (Narock and Lepping, 2007). We do not expect to find the same behaviour in all three components by the existence of anisotropy. An average value ($\langle D_{d1} \rangle$) of wavelet coefficient $D_{d1}$ in the three magnetic field components is calculated:

$$\langle D_{d1} \rangle = \frac{1}{3} \sum_{i=1}^{3} D_{d1}^{(i)}, \qquad (4)$$

where the angle brackets $\langle \cdots \rangle$ denote an average of the $D_{d1}$ in IMF components ( $i = 1, 2, 3 = B_x, B_y, B_z$ ). Its value is useful to compare the fluctuations between SW regions. From a physical point of view, this technique is useful to find candidate regions in the IMF dataset with more perturbations. The $\langle D_{d1} \rangle$ value increases with the degree of disorder and it is maximum for completely random systems.

The treatment procedure is able to characterize regular/non regular behaviour existing in experimental data to identify the transition between regions with these two primary behaviours in objective bases. The SW time interval is separated into three new time intervals (windows) corresponding to the preceding sheath or pre-MC, the MC itself, and the SW after the MC or post-MC.

The criterion to select a precise data window after the MC is empirical. Each post-MC

region was selected with the same length of the cloud regions. The main effort is to study SW data interval containing the ICMEs, where a shock event and a cloud region were reported. Arbitrary selection of post-MC region could affect the results, because this region could be disturbed by other processes unrelated with the MC itself. Thus, the physics of the system should not be changed in the proposed methodology. Further analyses of complicated events can indeed help to understand the true processes occurring in the interplanetary medium. In an evident way, showing the behaviour in the different regions is valuable because only then will be possible to justify that wavelet coefficients may help to find boundaries. A *zoom in* treatment in the fluctuations from variables with random variations (i.e., IMF) could help to separate disturbance processes, e.g., MC-candidate event inside of an ICME. Our hypothesis is that wavelet coefficients help to identify boundaries in the SW data, specifically the shock waves and the leading edge of ICMEs.

## 4. Results And Discussion

We present two case studies based on the analysis of Huttunen et al. (2005), where we have applied this methodology to analyse MC periods (events 14 and 16, Table 1). The study is extended to a total of 41 cases shown in the table, although the results are not presented individually here. In this section, a discussion is done to reach an interpretation.

### 4.1 February 11-13, 2000 ICME event

In Figure 1, at the top, we show the time series of IMF $B_z$ component measured by the ACE spacecraft at the date February 11; 23:23 UT-February 13; 12:00 UT; 2000. The data was measured in GSM coordinate system with resolution time of 16 s. The three regions under study are separated by two vertical dashed lines. At the bottom, we show the square of the first decomposition level of wavelet coefficients, d1, and results of $D_{d1}$. The mean of wavelet coefficient $D_{d1}$ in time series at plasma sheath is 0.828 nT$^2$. The result is that the lower $D_{d1}$ (0.156 nT$^2$) corresponds to the MC.

In Table 5, the results of $D_{d1}$ for the three components of $B$ are presented. Seen in the figure, the MC regions in the three components always have the lowest $D_{d1}$ value. While higher $D_{d1}$ values in all components correspond to the sheath region.

As a previously known feature, the larger amplitude of the wavelet coefficients, d1, are indeed associated with abrupt signal locally. From a visual inspection of data, detections may not be an easy task; but the wavelet transforms aids to find those kinds of phenomena.

### 4.2 July 11-14, 2000 ICME event

In Figure 2, a similar study is done. At the top, we show the time series of IMF $B_z$ component measured by ACE spacecraft at the date July 11; 11:22 UT-July 14; 05:00 UT; 2000. The three regions under study are separated by two vertical dashed lines. At the bottom, the square of first decomposition level of wavelet coefficient $d1^2$ versus time is plotted.

The statistical mean of the wavelet coefficient $D_{d1}$ in the sheath region is 0.625 nT$^2$. Again the lowest $D_{d1}$ (0.042 nT$^2$) corresponds to the MC region; and the higher $D_{d1}$ (0.625 nT$^2$) corresponds to the sheath region. The highest amplitude of $d1^2$ inside the third region (Post- MC) is due to the arrival of other event (event 17 in Table 1).

Related to this case, the results of $D_{d1}$ for the three components of $B$ are presented in Table 5. Also seen in the earlier figure, the MC region in the three components always has the lowest $D_{d1}$ value. While the highest $D_{d1}$ value in all components correspond to the sheath region.

The tendency of the MC events to have lower values of $D_{d1}$ in comparison with the processes of the other regions can be noticed. This feature is clearly identified by using this approach, which can be added to the usual features (Burlaga et al., 1981) established earlier for the MCs. Also, we found higher $D_{d1}$ values in the sheath. The higher amplitudes values of the wavelet coefficients indicate singularity patterns which are identified in the sheath region (see top panel on Figures 1 and 2).

## 4.3 41 ICMEs events

Aiming at a conclusive analysis, the calculations of $D_{d1}$ for the three IMF components are done for the other cases of Table 1. The procedure is identical to the one used in the previous studies.

In Figure 3, the $\langle D_{d1} \rangle$ values versus number of events were plotted respectively as squares, cross-circles symbols, and triangles symbols, correspond to the sheath, MC and Post-MC regions. We can compare the $\langle D_{d1} \rangle$ values of the three regions for every event. The $\langle D_{d1} \rangle$ values are higher in the sheath region in 35/41 or 85.4% events. This does not occur in the events numbered as 4; 5; 6; 13; 24; 34 in Table 1, where the highest values are found in the "Post-MC" regions. The explanation is that Post-MCs as shown in Figure 2, there may be an arrival of a shock or an ICMEs. However, the magnetic field fluctuation in the sheath is always greater than one in the cloud that follows. In particular, the magnetic field fluctuation in some MC regions (events numbered as 9; 19; 17; 20; 21; 31; 41) is greater than one in the SW that follows. We can learn that fluctuations are not low in all MCs, in some cases waves can penetrate from the sheath to the cloud. In this paper, the goal is to test the usefulness of this wavelet technique to study fluctuations in the SW data in order to explore any intrinsic physical process.

Figure 4 shows a histogram constructed from the occurrence frequencies of the $\langle D_{d1} \rangle$ values. The $\langle D_{d1} \rangle$ values for the sheath, MC and Post-MC regions are plotted respectively as grey, black, and white bins. In this figure, 63.4% of the MCs are located in the first two sets of bars on the left, while there are 4.9% and 24.4% of the sheaths and Post-MC regions respectively. The wavelet coefficients are low in some sheath regions. This means that if an ICME is not moving faster than the surrounding SW (Klein and Burlaga, 1982; Zhang and Burlaga, 1988; Burlaga, 1988), the sheath region does not present a very corrugated feature in the magnetic field. In principle, the identification by visual inspection could be more difficult to be done under these conditions. Conversely, in the last four sets of bars we have 75.6% of the sheaths and only 12.2% of the MCs regions. The results presented in the two previous case studies are confirmed: the largest amplitudes of the magnetic field fluctuations are in the sheath, and the lowest ones are during the MC. However, we do not have well defined $\langle D_{d1} \rangle$ values to identify the three different regions. Figure 4 only allows the comparison between values from the three regions in the same event. We can conclude that there is not a well-defined fluctuations pattern inside of MCs. The fluctuations could depend on the SW in the environment where the MC is expanding.

Figure 4 shows that due to the overlapping observed between the three distributions, this technique could not be used to identify boundaries automatically. It provides an objective analysis technique that helps in reducing the effort to find the boundaries inside of ICME, fundamentally the cloud boundaries. This technique could be useful to help a specialist to find boundaries when working with IMF dataset.

As in Table 5, the higher $D_{d1}$ values are found in $B_z$ component for every region. By direct visual inspection, most of the time this detection is not possible. However, the wavelet transform enables finding this phenomenon easily. The $B_z$ component is very important in the magnetic reconnection at Earth's magnetopause. An open question could be asked: how important are the fluctuations for the geoeffectiveness? We think that this is an important example of application of this technique in order to evaluate the SW fluctuations. Also, the wavelet coefficients can help to obtain a better visualization of the shock and to identify the initial border of the MC.

The wavelet coefficients recover the expected behaviours of the physical processes underlying the magnetic records. This is understandable, because the MC has a geometric structure in form of flux-rope, unlike the sheath region and the "quiet" SW. The sheath is naturally a turbulent region, presenting many fluctuations in the IMF data with large $D_{d1}$ values. A smoother magnetic field is the cause of the low values of $D_{d1}$ in the MCs regions. The existence of MCs with large values of the wavelet coefficients was an unexpected result in this study. We have found five MCs with this feature, and further they will deserve specific studies. The SW after the MC can present an extended quiet behaviour, or an increase of random characteristics, or even turbulences from an arrival event (for the latter, e.g., the events 16 and 20). Sometimes, the Post-MC region has a large $\langle D_{d1} \rangle$ value due to the existence of a reverse shock.

If this technique is applied to a large dataset of SW IMF, the wavelet coefficients could be also large in other regions in which there are no ICMEs. On other hand, the wavelet coefficients are relatively lower in quiet SW regions. Although it does not allow identifying clouds automatically, it is an useful tool for experts. Because, this technique can be used as auxiliary tools to find cloud boundaries, when, for example, the minimum variance analysis (MVA) is used. In fact, we have used for this purpose. In our opinion, the presentation of this tool for the Space Physics Community could "open doors" for other applications. For example, we believe that it might be useful to study Alfvén waves, where fluctuations in the SW with different pseudo-frequencies can be investigated.

## 4.4 Application to identify the shock and leader edge of ICME

The formation of a sheath implied in the existence of a shock waves. If we cannot find a shock then the sheath is not defined. However, if the MC is moving at the same speed as the ambient SW but still expanding, it will disturb both the SW ahead and behind, creating sheath-like structures (though they may not be bounded by a shock front). This study considers the events of MCs not associated with evident shock waves, presented in Table 3 (see event 3). With illustrative purpose, a case study is presented for the date June 24; 12:00 UT-June 25; 16:00 UT; 1998. The criteria to select the data interval after the MC are the same used previously. The duration time in regions at 41 sheaths is less than one day, and then a region with this equivalent duration from the initial time of the cloud is chosen.

In Figure 5, the above interval at the date June 23; 12:00 UT-June 26; 16:00 UT; 1998 is shown. Each panel presents respectively, from top to bottom, $B_x$, $B_y$ and $B_z$ time series respectively. At the bottom of the respective panels, the square of the first decomposition level of wavelet coefficients, d1, versus time is plotted. The two vertical dashed lines correspond to the MC region delimitations identified by Huttunen et al. (2005). The wavelet coefficients allow for a *zoom in* on the fluctuations of magnetic components. As larger amplitudes in the wavelet coefficients are observed inside the initial border of MC, then we think that this boundary should be redefined. So, the leader edge at date June 24; 16:32 UT 1998 is redefined. The second vertical thick line corresponds to the previous data.

Also, wavelet coefficients could be used to identify sheath like structures. However, the confirmation on the type of electrodynamical discontinuity implies the use of plasma data. So, a probable discontinuity at date June 24; 04:00 UT 1998 was identified. Thus, with the help of SW plasma parameters, an interplanetary sheath-like structure can be associated to this event. The first vertical thick line corresponds to the start of its location.

In Figure 5 (all panels), the $D_{d1}$ values in each regions are shown. We found higher $D_{d1}$ values in the sheath-like structures while the lower values correspond to cloud region. The results related to this part are consistent with the earlier results.

In conclusion, this methodology has a practical application. Maybe other applications for Space Physics Community uses will be found, mainly taking into account fluctuations that occur in several frequency ranges.

## 5. Conclusions

We deal with time series of SW for a group of magnetic clouds in order to analyse the fluctuations of the IMF $B_x$, $B_y$ and $B_z$ components. The mathematical property chosen here was the statistical mean of the wavelet coefficients ($\langle D_{d1} \rangle$) which was obtained by applying the discrete orthogonal wavelet transform using Daubechies wavelet of order two (i.e. Daubechies scale filters order 2, db2) to the components of the IMF as recorded by the instruments of the MAG on-board of the ACE S/C at the L1 point.

The main point in the use of the amplitude of the Daubechies wavelet coefficients is that they represent the local regularity present in the signal in study (Mallat, 1989). They were constructed to express the local approximation error between a certain local polynomial reproduction and the signal itself. This is used to identify local regularity in high order derivatives in the analysed signal. The local regularity changes can be therefore highlighted by means of the amplitude wavelet coefficients. It is not easy or even possible to see discontinuities in high order derivatives that cause disturbances by visual inspection of the signal. For instance, using Daubechies wavelet of order 2, discontinuities higher than the first derivatives can be detected and measured, respectively. We use that propriety of local regularity identification to highlight possible regions of regularity on the magnetic field at three different regions around an ICME event measure at IMF datasets. The results show that there is, apparently, a clear distinction between the values of the wavelet coefficients obtained along the different parts of the passing magnetic structure (ahead of the MC, i.e., the sheath; the MC itself; and after the passage of the MC (Post-MC)). The measurements show that $\langle D_{d1} \rangle$ exhibits the lower values during the passage of the MC. Also, we found higher values in the sheaths.

Using assumptions that concern the physics of MC, the analyses developed in this work show that a smoothed magnetic configuration (i.e., few magnetic fluctuations) in MC is the main reason of the lower values of wavelet coefficients during it. This study has been performed only for specific types of ICMEs, all of which were structures that appeared to be MCs. This tool allows for the comparison of the existing fluctuation of SW magnetic field, i.e., $B_x$, $B_y$, and $B_z$, which it is not an easy task under simple visual inspection. The $B_x$ component has lower fluctuations, or singularities, and the $B_z$ component the higher ones.

We can identify the effect of shock waves in the change of the local regularity of the IMF component using its $d1^2$ time series, shows that the amplitude of the wavelet coefficients decreases at transient regions in MC boundaries identified by other authors. Therefore, the behaviour expected inside of MCs is the decrease of entropy and variance respectively, and then the fluctuations should be lower than outside them. The previous behaviour is not true for all the cases because some another phenomenon could also be present. However, in this study this was verified for 32/41 or 78% of the cases. We can learn that fluctuations are not low in all magnetic clouds, in some cases waves can penetrate from the sheath to the cloud. The fluctuations could depend on the solar wind in the environment where the cloud is expanding.

This is an objective analysis technique provided to find the boundaries of magnetic clouds related to ICMEs. The procedure identifies transitions in the IMF regularity for different regions existing in the solar wind, which highlight cloud regions. It can be very useful for specialists, because the wavelet coefficients have the advantage to find some discontinuities (transients) that are not easy to be seen in the IMF data by simple visual inspection.

By now, only assumptions for proper MCs were validated. Maybe this methodology could be extended to identify features of some other specific fluctuation patterns in the IMF, such as CIR, heliospheric current sheath crossings or ICMEs without MC signatures which has not yet been done.. Such an approach aiming at new facilities for the Space Physics community efforts seem to be an important contribution.

## . 6 Acknowledgments

This work was supported by grants from CNPq (grants 483226 /2011-4, 307511 /2010-3, 306828 /2010-3 and 486165 /2006-0), FAPESP (grants 2012 /072812-2, and 2007 /07723-7) and CAPES (grants 1236-83 /2012 and 86 /2010-29). A.O.G. thanks the CAPES and CNPq (grant 141549/2010-6) for his PhD scholarship and CNPq (grant 150595/2013-1) for his postdoctoral research support. V. E. Menconi thanks to the grants FAPESP 2008/09736-1, CNPq 3124862012-0 and 455097/2013-5. We also wish to thank the anonymous referees for improvement of this paper.

## Appendix A. The wavelet coefficients in a discontinuous function.

The local regularity changes can be therefore highlighted by means of the amplitude wavelet coefficients. Using the signal presented in (Daubechies, 1992, p.301), we constructed the following example to illustrate the propriety.

Considering,

$$f(x) = \begin{cases} 2e^{-|x|} & if \quad x \leq -1, \\ e^{-|x|} & if \quad -1 < x \leq 1, \\ e^{-x}[(x-1)^2 + 1] & if \quad x > 1; \end{cases}$$

This function presents a discontinuity at x =, -1, a discontinuity in the first derivative at x= 0 and a discontinuity in the second derivative at x= 1. We have computed decomposition level of discrete orthogonal wavelet transform using a Daubechies wavelet of order 2. The result is presented in the Fig. A.6, the larger amplitude of the wavelet coefficients, identifies the three points where the signal has changes in the local regularity.

Table 1: Solar wind data studied (from Huttunen et al. (2005)).

| No. | Year | Shock, UT | MC start, UT | MC stop, UT | Post-MC, UT |
| --- | --- | --- | --- | --- | --- |
| 01 | 1998 | 06 Jan, 13:19 | 07 Jan, 03:00 | 08 Jan, 09:00 | 10 Jan, 15:00 |
| 02 |  | 03 Feb, 13:09 | 04 Feb, 05:00 | 05 Feb, 14:00 | 06 Feb, 23:00 |
| 03 |  | 04 Mar, 11:03 | 04 Mar, 15:00 | 05 Mar, 21:00 | 07 Mar, 03:00 |
| 04 |  | 01 May, 21:11 | 02 May, 12:00 | 03 May, 17:00 | 04 May, 22:00 |
| 05 |  | 13 Jun, 18:25 | 14 Jun, 02:00 | 14 Jun, 24:00 | 15 Jun, 22:00 |
| 06 |  | 19 Aug, 05:30 | 20 Aug, 08:00 | 21 Aug, 18:00 | 23 Aug, 04:00 |
| 07 |  | 24 Sep, 23:15 | 25 Sep, 08:00 | 26 Sep, 12:00 | 27 Sep, 16:00 |
| 08 |  | 18 Oct, 19:00 | 19 Oct, 04:00 | 20 Oct, 06:00 | 21 Oct, 08:00 |
| 09 |  | 08 Nov, 04:20 | 08 Nov, 23:00 | 10 Nov, 01:00 | 12 Nov, 02:00 |
| 10 |  | 13 Nov, 00:53 | 13 Nov, 04:00 | 14 Nov, 06:00 | 15 Nov, 08:00 |
| 11 | 1999 | 18 Feb, 02:08 | 18 Feb, 14:00 | 19 Feb, 11:00 | 20 Feb, 08:00 |
| 12 |  | 16 Apr, 10:47 | 16 Apr, 20:00 | 17 Apr, 18:00 | 18 Apr, 16:00 |
| 13 |  | 08 Aug, 17:45 | 09 Aug, 10:00 | 10 Aug, 14:00 | 11 Aug, 18:00 |
| 14 | 2000 | 11 Feb, 23:23 | 12 Feb, 12:00 | 12 Feb, 24:00 | 13 Feb, 12:00 |
| 15 |  | 20 Feb, 20:57 | 21 Feb, 14:00 | 22 Feb, 12:00 | 23 Feb, 10:00 |
| 16 |  | 11 Jul, 11:22 | 11 Jul, 23:00 | 13 Jul, 02:00 | 14 Jul, 05:00 |
| 17 |  | 13 Jul, 09:11 | 13 Jul, 15:00 | 13 Jul, 24:00 | 14 Jul, 09:00 |
| 18 |  | 15 Jul, 14:18 | 15 Jul, 19:00 | 16 Jul, 12:00 | 17 Jul, 05:00 |
| 19 |  | 28 Jul, 05:53 | 28 Jul, 18:00 | 29 Jul, 10:00 | 30 Jul, 02:00 |
| 20 |  | 10 Aug, 04:07 | 10 Aug, 20:00 | 11 Aug, 08:00 | 11 Aug, 20:00 |
| 21 |  | 11 Aug, 18:19 | 12 Aug, 05:00 | 13 Aug, 02:00 | 13 Aug, 23:00 |
| 22 |  | 17 Sep, 17:00 | 17 Sep, 23:00 | 18 Sep, 14:00 | 19 Sep, 05:00 |
| 23 |  | 02 Oct, 23:58 | 03 Oct, 15:00 | 04 Oct, 14:00 | 05 Oct, 13:00 |
| 24 |  | 02 Oct, 23:58 | 13 Oct, 17:00 | 14 Oct, 13:00 | 15 Oct, 09:00 |
| 25 |  | 28 Oct, 09:01 | 28 Oct, 24:00 | 29 Oct, 23:00 | 30 Oct, 22:00 |
| 26 |  | 06 Nov, 09:08 | 06 Nov, 22:00 | 07 Nov, 15:00 | 08 Nov, 08:00 |
| 27 | 2001 | 19 Mar, 10:12 | 19 Mar, 22:00 | 21 Mar, 23:00 | 23 Mar, 24:00 |
| 28 |  | 27 Mar, 17:02 | 27 Mar, 22:00 | 28 Mar, 05:00 | 28 Mar, 12:00 |
| 29 |  | 11 Apr, 15:18 | 12 Apr, 10:00 | 13 Apr, 06:00 | 14 Apr, 02:00 |
| 30 |  | 21 Apr, 15:06 | 21 Apr, 23:00 | 22 Apr, 24:00 | 24 Apr, 01:00 |
| 31 |  | 28 Apr, 04:31 | 28 Apr, 24:00 | 29 Apr, 13:00 | 30 Apr, 02:00 |
| 32 |  | 27 May, 14:17 | 28 May, 11:00 | 29 May, 06:00 | 30 May, 01:00 |
| 33 |  | 31 Oct, 12:53 | 31 Oct, 22:00 | 02 Nov, 04:00 | 03 Nov, 10:00 |
| 34 | 2002 | 23 Mar, 10:53 | 24 Mar, 10:00 | 25 Mar, 12:00 | 26 Mar, 14:00 |
| 35 |  | 17 Apr, 10:20 | 17 Apr, 24:00 | 19 Apr, 01:00 | 20 Apr, 02:00 |
| 36 |  | 18 May, 19:44 | 19 May, 04:00 | 19 May, 22:00 | 20 May, 16:00 |

| 37 |      | 01 Aug, 23:10 | 02 Aug, 06:00 | 02 Aug, 22:00 | 03 Aug, 14:00 |
| 38 |      | 30 Sep, 07:55 | 30 Sep, 23:00 | 01 Oct, 15:00 | 02 Oct, 07:00 |
| 39 | 2003 | 20 Mar, 04:20 | 20 Mar, 13:00 | 20 Mar, 22:00 | 21 Mar, 07:00 |
| 40 |      | 17 Aug, 13:41 | 18 Aug, 06:00 | 19 Aug, 11:00 | 20 Aug, 16:00 |
| 41 |      | 20 Nov, 07:27 | 20 Nov, 11:00 | 21 Nov, 01:00 | 22 Nov, 15:00 |

**Table 2:** MC events measured by WIND (not examined). Letter "Q" denotes whether the event was an MC (l) or cloud candidate (cl).

| No. | Year | Shock, UT | MC start, UT | MC stop, UT | Q |
|---|---|---|---|---|---|
| 01 | 1997 | 10 Jan, 00:20 | 10 Jan, 05:00 | 11 Jan, 02:00 | l |
| 02 |      | 09 Feb, 23:43 | 10 Feb, 03:00 | 10 Feb, 19:00 | cl |
| 03 |      | 10 Apr, 12:57 | 11 Apr, 08:00 | 11 Apr, 16:00 | l |
| 04 |      | -             | 21 Apr, 17:00 | 22 Apr, 24:00 | cl |
| 05 |      | 15 May, 00:56 | 15 May, 10:00 | 15 May, 24:00 | l |
| 06 |      | -             | 15 May, 07:00 | 16 May, 16:00 | l |
| 07 |      | 26 May, 09:10 | 26 May, 16:00 | 27 May, 19:00 | l |
| 08 |      | -             | 09 Jun, 06:00 | 09 Jun, 23:00 | l |
| 09 |      | 19 Jun, 00:12 | 19 Jun, 06:00 | 19 Jun, 16:00 | l |
| 10 |      | -             | 15 Jul, 09:00 | 16 Jul, 06:00 | l |
| 11 |      | -             | 03 Aug, 14:00 | 04 Aug, 02:00 | l |
| 12 |      | -             | 18 Sep, 03:00 | 19 Sep, 21:00 | l |
| 13 |      | -             | 22 Sep, 01:00 | 22 Sep, 18:00 | l |
| 14 |      | 01 Oct, 00:20 | 01 Oct, 15:00 | 02 Oct, 22:00 | l |
| 15 |      | 10 Oct, 15:48 | 10 Oct, 23:00 | 12 Oct, 01:00 | l |
| 16 |      | 06 Nov, 22:07 | 07 Nov, 05:00 | 08 Nov, 03:00 | l |
| 17 |      | 22 Nov, 08:55 | 22 Nov, 19:00 | 23 Nov, 12:00 | l |

**Table 3:** These magnetic cloud events are not preceded by shock waves. Letter "Q" denotes whether the event was an MC (l) or cloud candidate (cl).

| No. | Year | Shock, UT | MC start, UT | MC stop, UT | Q |
|---|---|---|---|---|---|
| 01 | 1998 | - | 17 Feb, 10:00 | 18 Feb, 04:00 | l |
| 02 | | - | 02 Jun, 10:00 | 02 Jun, 16:00 | l |
| 03 | | - | 24 Jun, 12:00 | 25 Jun, 16:00 | l |
| 04 | 1999 | - | 25 Mar, 16:00 | 25 Mar, 23:00 | l |
| 05 | | - | 21 Apr, 12:00 | 22 Apr, 13:00 | l |
| 06 | | - | 22 Aug, 12:00 | 23 Aug, 06:00 | l |
| 07 | | - | 21 Sep, 20:00 | 23 Sep, 05:00 | l |
| 08 | | - | 14 Nov, 01:00 | 14 Nov, 09:00 | cl |
| 09 | | - | 16 Nov, 09:00 | 16 Nov, 23:00 | l |
| 10 | 2000 | - | 15 Jul, 05:00 | 15 Jul, 14:00 | cl |
| 11 | | - | 31 Jul, 22:00 | 01 Aug, 12:00 | l |
| 12 | 2001 | - | 04 Mar, 16:00 | 05 Mar, 01:00 | l |
| 13 | | - | 18 Jun, 23:00 | 19 Jun, 14:00 | l |
| 14 | | - | 10 Jul, 17:00 | 11 Jul, 23:00 | l |
| 15 | | 03 Oct, 08:?? | 03 Oct, 01:00 | 03 Oct, 16:00 | l |
| 16 | | - | 24 Nov, 17:00 | 25 Nov, 13:00 | cl |
| 17 | 2002 | - | 28 Feb, 18:00 | 01 Mar, 10:00 | l |
| 18 | | - | 19 Mar, 22:00 | 20 Mar, 10:00 | l |
| 19 | | - | 20 Apr, 13:00 | 21 Apr, 15:00 | l |
| 20 | | 23 May, 10:15 | 23 May, 22:00 | 24 May, ??:?? | cl |
| 21 | 2003 | - | 27 Jan, 01:00 | 24 May, ??:?? | l |
| 22 | | - | 29 Oct, 12:00 | 30 Oct, 01:00 | l |

**Table 4:** Pseudo-period (seconds) regarding the Daubechies orthogonal wavelets. In this work $\Lambda = 16\,s$, j = 1 and db2 then pseudo-period is 48 s. The information here could be useful for studying fluctuations with different frequencies.

| Level | Order | | | |
|---|---|---|---|---|
| j | 1 | 2 | 3 | 4 |
| 1 | 32.1 | 48.0 | 40.0 | 44.8 |
| 2 | 64.3 | 96.0 | 80.0 | 89.6 |
| 3 | 128.5 | 192.0 | 160.0 | 179.2 |
| 4 | 257.0 | 384.0 | 320.0 | 358.4 |
| 5 | 514.0 | 768.0 | 640.0 | 716.8 |

**Table 5:** Mean $D_{d1}$ of wavelet coefficients.

| Events | $D_{d1}\,B_x$ | $D_{d1}\,B_y$ | $D_{d1}\,B_z$ | $\langle D_{d1} \rangle$ |
|---|---|---|---|---|
| *Feb 11-13, 2000:* | | | | |
| *Sheath* | 0.524 | 0.814 | 0.828 | 0.722 |
| *MC* | 0.093 | 0.124 | 0.156 | 0.124 |
| *Post-MC* | 0.177 | 0.247 | 0.319 | 0.248 |
| **Jul 11-14, 2000:** | | | | |
| *Sheath* | 0.279 | 0.270 | 0.625 | 0.391 |
| *MC* | 0.016 | 0.032 | 0.042 | 0.030 |
| *Post-MC* | 0.233 | 0.230 | 0.458 | 0.307 |

**Figure 1:** At the top, IMF $B_z$ (in GSM system) versus time from the ACE spacecraft with 16s time resolution, at February 11; 23:23 UT-February 13; 12:00 UT; 2000. At the bottom, the square of the first decomposition level of wavelet coefficient $d1^2$ versus time for the sheath region (left of the first vertical dashed line), the MC (middle between the vertical dashed lines), and the quiet SW (right of the second vertical dashed line). The lower values of $D_{d1}$ are noticed inside of MC region.

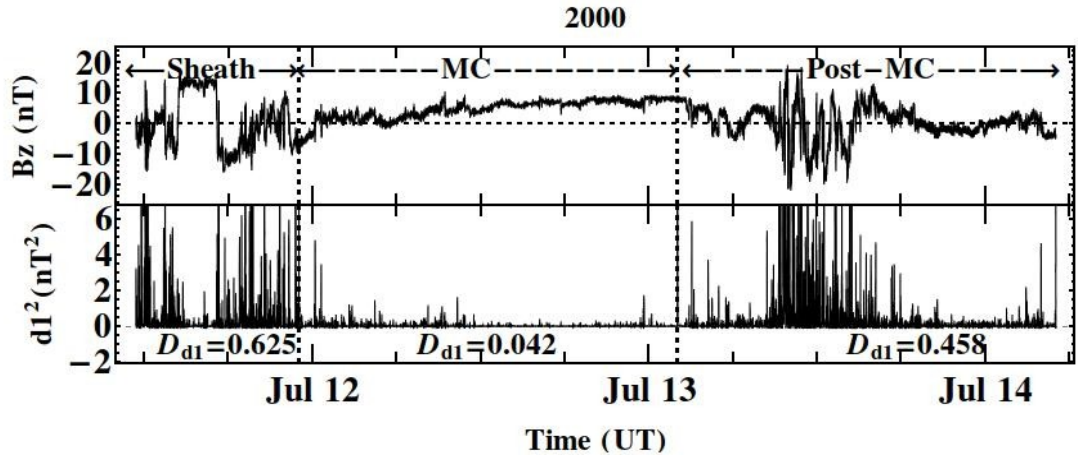

**Figure 2:** (At the top, IMF $B_z$ (in GSM system) versus time from the ACE spacecraft with 16 s time resolution, at July 11; 11:22 UT-July 14; 05:00 UT; 2000. At the bottom, the square of the first decomposition level of wavelet coefficient $d1^2$ versus time for the sheath region (left of the first vertical dashed line), the MC (middle between the vertical dashed lines), and the quiet SW (right of the second vertical dashed line). The high amplitude of $d1^2$ inside the third region (Post-MC) is because other event arrived. The lower values of $D_{d1}$ is noticed inside of MC region.

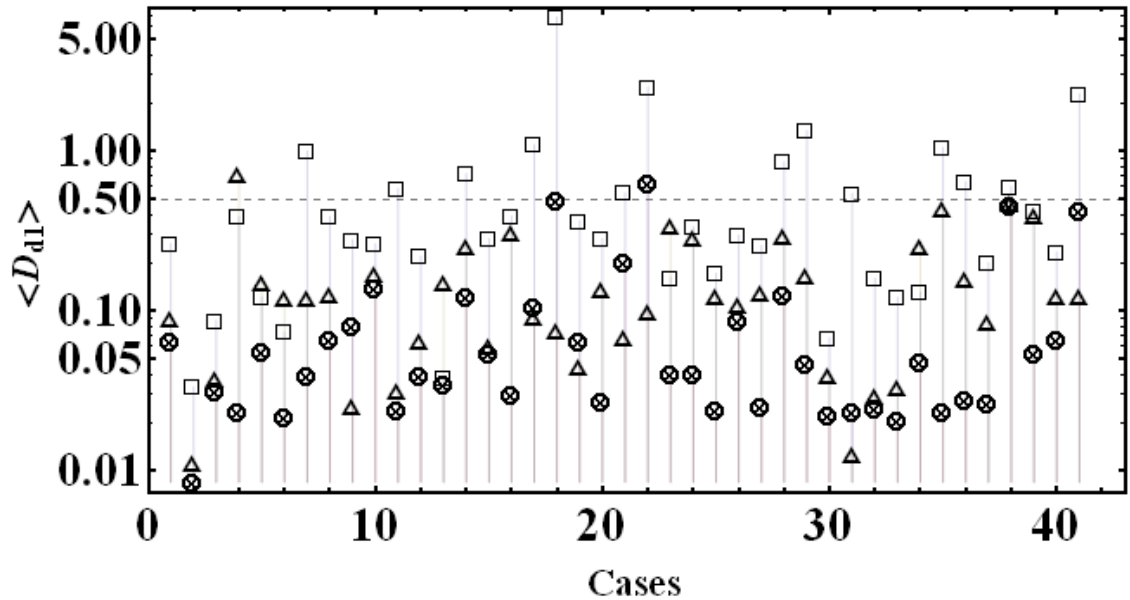

**Figure 3:** The $\langle D_{d1} \rangle$ values versus number of events were plotted respectively as squares, cross-circles symbols, and triangles symbols, correspond to the sheath, MC and Post-MC regions. The y axis is plot with a logarithmic scale, because is best to visualization.

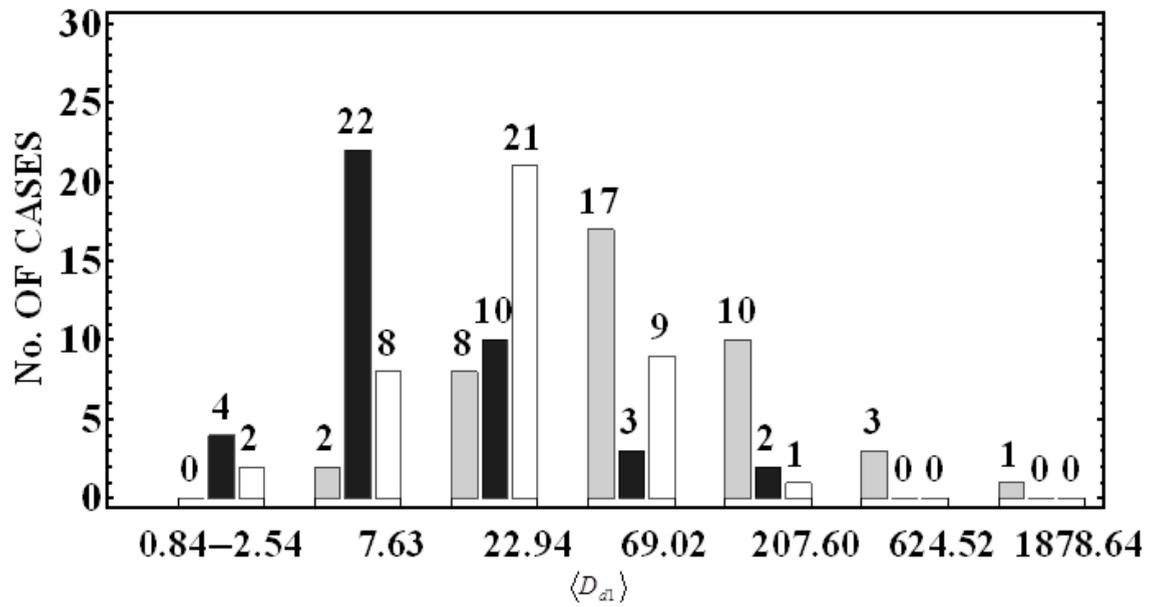

**Figure 4:** A histogram is constructed from a frequency table of $\langle D_{d1} \rangle$ values; the abscissa axis was normalized by 0.01. The $\langle D_{d1} \rangle$ values for the sheath, MC and post-MC, the three select regions, plotted as the grey, black and white.

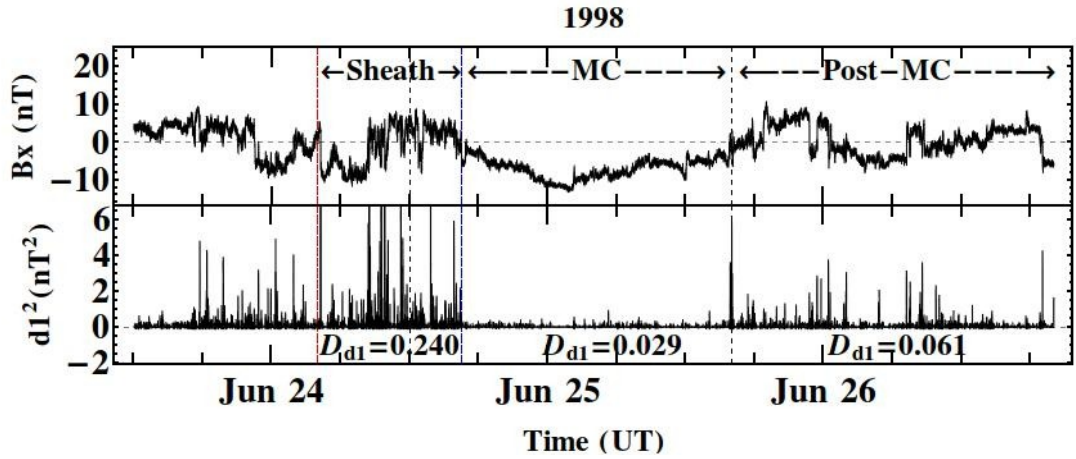

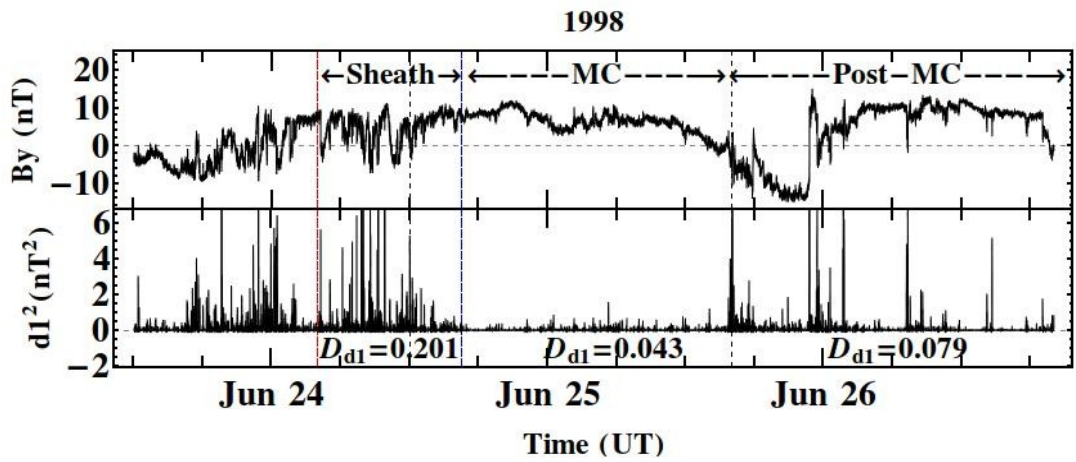

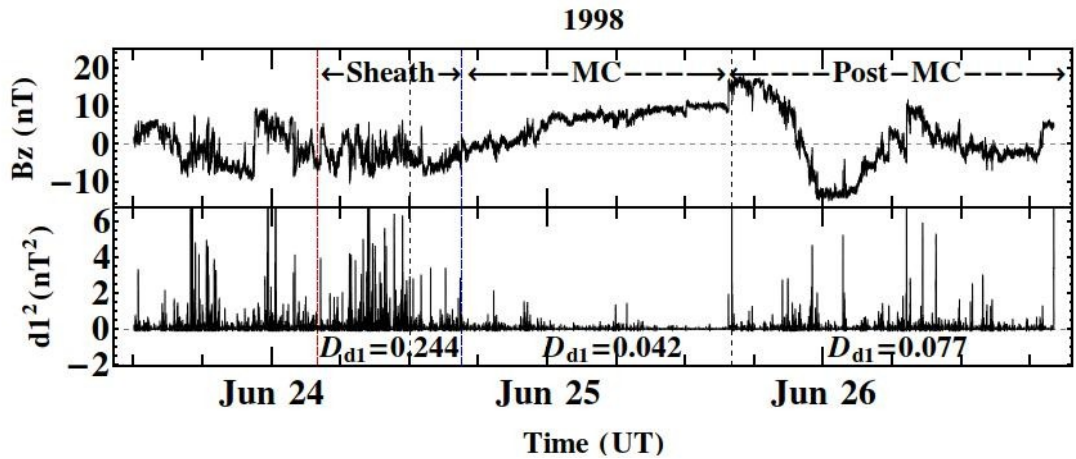

**Figure 5:** (top the panel) At top, IMF $B_x$ (in GSM system) versus time from the ACE spacecraft with 16 s time resolution, at date June 23; 12:00 UT-June 26; 16:00 UT; 1998; at the bottom, the square of the first decomposition level of wavelet coefficient $d1^2$ versus time. Also, the other two components must be analysed, as is shown in the middle and bottom panels.

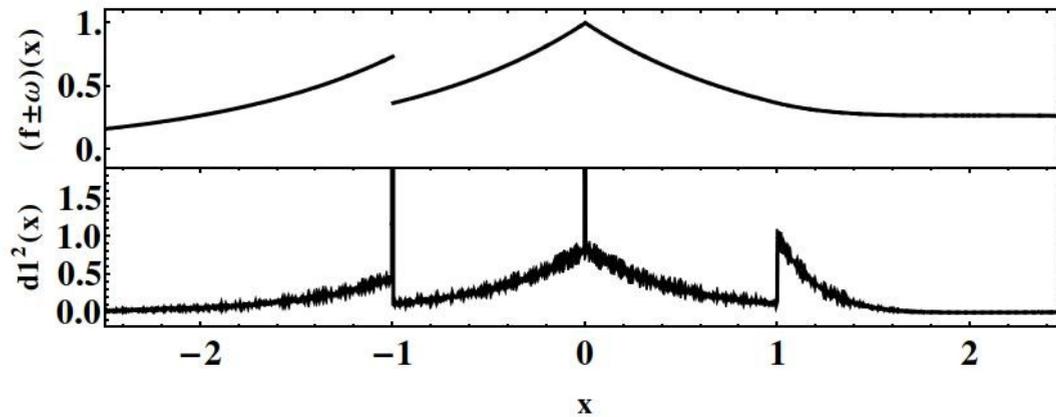

Figure A.6: At the top, the signal $f(x) \pm \omega(x)$ versus x was plotted, where $\omega(x)$ is a white noise. At the bottom, the square of the first decomposition level of wavelet coefficients $d1^2(x)/10^{-12}$ versus x was plotted.